\documentstyle[twoside]{article}

\catcode`\@=11
\long\def\@makefntext#1{
\protect\noindent \hbox to 3.2pt {\hskip-.9pt
$^{{\eightrm\@thefnmark}}$\hfil}#1\hfill}		%CAN BE USED

\def\thefootnote{\fnsymbol{footnote}}
\def\@makefnmark{\hbox to 0pt{$^{\@thefnmark}$\hss}}	%ORIGINAL

\def\ps@myheadings{\let\@mkboth\@gobbletwo
\def\@oddhead{\hbox{}
\rightmark\hfil\eightrm\thepage}
\def\@oddfoot{}\def\@evenhead{\eightrm\thepage\hfil
\leftmark\hbox{}}\def\@evenfoot{}
\def\sectionmark##1{}\def\subsectionmark##1{}}

\oddsidemargin=\evensidemargin
\renewcommand{\thefootnote}{\fnsymbol{footnote}}

\newcounter{sectionc}
\newcounter{subsectionc}
\newcounter{subsubsectionc}
\renewcommand{\section}[1] {\vspace{12pt}\addtocounter{sectionc}{1}
\setcounter{subsectionc}{0}\setcounter{subsubsectionc}{0}\noindent
	{\tenbf\thesectionc. #1}\par\vspace{5pt}}
\renewcommand{\subsection}[1] {\vspace{12pt}
\addtocounter{subsectionc}{1}\setcounter{subsubsectionc}{0}\noindent
	{\bf\thesectionc.\thesubsectionc.
        {\kern1pt \bfit #1}}\par\vspace{5pt}}
\renewcommand{\subsubsection}[1] {\vspace{12pt}
\addtocounter{subsubsectionc}{1}\noindent
        {\tenrm\thesectionc.\thesubsectionc.\thesubsubsectionc.
	{\kern1pt \tenit #1}}\par\vspace{5pt}}

\newcounter{appendixc}
\newcounter{subappendixc}[appendixc]
\newcounter{subsubappendixc}[subappendixc]
\renewcommand{\thesubappendixc}{\Alph{appendixc}.
        \arabic{subappendixc}}
\renewcommand{\thesubsubappendixc}{\Alph{appendixc}.
        \arabic{subappendixc}.\arabic{subsubappendixc}}

\renewcommand{\appendix}[1] {\vspace{12pt}
        \refstepcounter{appendixc}
        \setcounter{figure}{0}
        \setcounter{table}{0}
        \setcounter{lemma}{0}
        \setcounter{theorem}{0}
        \setcounter{corollary}{0}
        \setcounter{definition}{0}
        \setcounter{equation}{0}
        \renewcommand{\thefigure}{\Alph{appendixc}.\arabic{figure}}
        \renewcommand{\thetable}{\Alph{appendixc}.\arabic{table}}
        \renewcommand{\theappendixc}{\Alph{appendixc}}
        \renewcommand{\thelemma}{\Alph{appendixc}.\arabic{lemma}}
        \renewcommand{\thetheorem}{\Alph{appendixc}.\arabic{theorem}}
        \renewcommand{\thedefinition}{\Alph{appendixc}.
         \arabic{definition}}
        \renewcommand{\thecorollary}{\Alph{appendixc}.
         \arabic{corollary}}
        \renewcommand{\theequation}{\Alph{appendixc}.
         \arabic{equation}}
        \noindent{\tenbf Appendix \theappendixc #1}\par\vspace{5pt}}
\newcommand{\subappendix}[1] {\vspace{12pt}
        \refstepcounter{subappendixc}
        \noindent{\bf Appendix \thesubappendixc. {\kern1pt \bfit #1}}
	\par\vspace{5pt}}
\newcommand{\subsubappendix}[1] {\vspace{12pt}
        \refstepcounter{subsubappendixc}
        \noindent{\rm Appendix \thesubsubappendixc.
        {\kern1pt \tenit #1}}\par\vspace{5pt}}

\newcommand{\textlineskip}{\baselineskip=13pt}
\newcommand{\smalllineskip}{\baselineskip=10pt}

\def\eightcirc{
\begin{picture}(0,0)
\put(4.4,1.8){\circle{6.5}}
\end{picture}}
\def\eightcopyright{\eightcirc\kern2.7pt\hbox{\eightrm c}}

\def\abstracts#1#2#3{{
	\centering{\begin{minipage}{4.5in}\baselineskip=10pt
        \footnotesize
	\parindent=0pt #1\par
	\parindent=15pt #2\par
	\parindent=15pt #3
	\end{minipage}}\par}}

\renewenvironment{thebibliography}[1]
	{\frenchspacing
	 \ninerm\baselineskip=11pt
	 \begin{list}{\arabic{enumi}.}
	{\usecounter{enumi}\setlength{\parsep}{0pt}
	 \setlength{\leftmargin 12.7pt}{\rightmargin 0pt}
	 \setlength{\itemsep}{0pt} \settowidth
	{\labelwidth}{#1.}\sloppy}}{\end{list}}

\newcounter{itemlistc}
\newcounter{romanlistc}
\newcounter{alphlistc}
\newcounter{arabiclistc}

\newcommand{\fcaption}[1]{
        \refstepcounter{figure}
        \setbox\@tempboxa = \hbox{\footnotesize Fig.~\thefigure. #1}
        \ifdim \wd\@tempboxa > 5in
           {\begin{center}
        \parbox{5in}{\footnotesize\smalllineskip Fig.~\thefigure. #1}
            \end{center}}
        \else
             {\begin{center}
             {\footnotesize Fig.~\thefigure. #1}
              \end{center}}
        \fi}

\newcommand{\tcaption}[1]{
        \refstepcounter{table}
        \setbox\@tempboxa = \hbox{\footnotesize Table~\thetable. #1}
        \ifdim \wd\@tempboxa > 5in
           {\begin{center}
        \parbox{5in}{\footnotesize\smalllineskip Table~\thetable. #1}
            \end{center}}
        \else
             {\begin{center}
             {\footnotesize Table~\thetable. #1}
              \end{center}}
        \fi}

\def\@citex[#1]#2{\if@filesw\immediate\write\@auxout
	{\string\citation{#2}}\fi
\def\@citea{}\@cite{\@for\@citeb:=#2\do
	{\@citea\def\@citea{,}\@ifundefined
	{b@\@citeb}{{\bf ?}\@warning
	{Citation `\@citeb' on page \thepage \space undefined}}
	{\csname b@\@citeb\endcsname}}}{#1}}

\newif\if@cghi
\def\cite{\@cghitrue\@ifnextchar [{\@tempswatrue
	\@citex}{\@tempswafalse\@citex[]}}
\def\citelow{\@cghifalse\@ifnextchar [{\@tempswatrue
	\@citex}{\@tempswafalse\@citex[]}}
\def\@cite#1#2{{$\null^{#1}$\if@tempswa\typeout
	{IJCGA warning: optional citation argument
	ignored: `#2'} \fi}}

\def\pmb#1{\setbox0=\hbox{#1}
	\kern-.025em\copy0\kern-\wd0
	\kern.05em\copy0\kern-\wd0
	\kern-.025em\raise.0433em\box0}

\def\fnt#1#2{\footnotetext{\kern-.3em
	{$^{\mbox{\scriptsize #1}}$}{#2}}}

\def\fpage#1{\begingroup
\voffset=.3in
\thispagestyle{empty}\begin{table}[b]\centerline{\footnotesize #1}
	\end{table}\endgroup}

\font\tenrm=cmr10
\font\tenit=cmti10
\font\tenbf=cmbx10
\font\bfit=cmbxti10 at 10pt
\font\ninerm=cmr9

\font\eightrm=cmr8

\def\qed{\hbox{${\vcenter{\vbox{		%HOLLOW SQUARE
   \hrule height 0.4pt\hbox{\vrule width 0.4pt height 6pt
   \kern5pt\vrule width 0.4pt}\hrule height 0.4pt}}}$}}

\renewcommand{\thefootnote}{\fnsymbol{footnote}}
\input epsf
\global\arraycolsep=2pt
\def\spose#1{\hbox to 0pt{#1\hss}}
\def\lsim{\mathrel{\spose{\lower 3pt\hbox{$\mathchar"218$}}
 \raise 2.0pt\hbox{$\mathchar"13C$}}}
\def\gsim{\mathrel{\spose{\lower 3pt\hbox{$\mathchar"218$}}
 \raise 2.0pt\hbox{$\mathchar"13E$}}}

\renewcommand{\theequation}{\thesection.\arabic{equation}}
\def\laq{~\raise 0.4ex\hbox{$<$}\kern -0.8em\lower 0.62
ex\hbox{$\sim$}~}
\def\gaq{~\raise 0.4ex\hbox{$>$}\kern -0.7em\lower 0.62
ex\hbox{$\sim$}~}

\def\beq{\begin{equation}}
\def\eeq{\end{equation}}
\def\bea{\begin{eqnarray}}
\def\eea{\end{eqnarray}}

\def \pa {\partial}
\def \ra {\rightarrow}

\def \ls {\lambda_s}
\def \La {\Lambda}

\def \b {\beta}
\def \a {\alpha}
\def \ap {\alpha^{\prime}}

\def \ga {\gamma}
\def \sg {\sigma}
\def \da {\delta}
\def \ep {\epsilon}
\def \r {\rho}

\def \Om {\Omega}
\def \noi {\noindent}

\def \hp {\widehat \phi}
\def \hpd {{\dot{\hp}}}

\begin{document}

%%% start CERN preprint title page %%%%%%%%%%%%%
\begin{titlepage}

\begin{flushright}
BA-TH/03-473\\
hep-th/0310293
\end{flushright}

\vspace{3 cm}

\begin{center}
\Large\bf Late-time effects of Planck-scale cosmology:
dilatonic interpretation of the dark energy field
\end{center}

\vspace{2cm}

\begin{center}
M. Gasperini\\
\vspace{0.3cm}
{\sl Dipartimento di Fisica, Universit\`a di Bari,}\\
{\sl Via Amendola 173, 70126 Bari, Italy}\\
and\\
{\sl Istituto Nazionale di Fisica Nucleare, Sezione di Bari, Italy}
\end{center}

\vspace{2cm}

\begin{abstract}

We present a model of dark energy based on the string effective
action, and on the assumption that the dilaton is strongly coupled to
dark matter. We discuss the main differences between this
class of models and  more conventional models of quintessence,
uncoupled to dark matter. This paper is based on talks presented at the
{\sl ``VII Congresso Nazionale di  Cosmologia"} 
(Osservatorio Astronomico di Roma, Monte Porzio Catone, November 
2002), and at the  Meeting {\sl ``Dark Energy Day"} 
(University of Milano-Bicocca, November 2002). To appear in the Proc. 
of  the International Conference on {\sl ``Thinking, Observing and Mining
the Universe"}  (Sorrento, September 2003), eds. G. Longo and G. Miele 
(World Scientific, Singapore).   
\end{abstract}

\vfill

\end{titlepage}

%\thispagestyle{empty}
%\vbox{}
%\newpage
%%% end CERN preprint title page %%%%%%%%%%%%%

\normalsize\textlineskip
\thispagestyle{empty}
\setcounter{page}{1}

%\copyrightheading{}			%{Vol. 0, No. 0 (1993) 000--000}

\fpage{1}

\centerline{\bf LATE-TIME EFFECTS OF PLANCK-SCALE COSMOLOGY: }

\centerline{\bf DILATONIC INTERPRETATION OF THE DARK ENERGY FIELD}

\vspace*{0.27truein}

\centerline{\footnotesize MAURIZIO GASPERINI}
\vspace*{0.015truein}
\centerline{\footnotesize\it Dipartimento di Fisica,  
Universit\`a di Bari,}
\baselineskip=10pt
\centerline{\footnotesize\it and Istituto Nazionale di Fisica Nucleare,
Sezione di Bari,}
\baselineskip=10pt
\centerline{\footnotesize  {\it Via Amendola 173, 70126 Bari, Italy}}
\baselineskip=10pt

\vspace*{0.25truein}
\abstracts
{We present a model of dark energy based on the string effective
action, and on the assumption that the dilaton is strongly coupled to
dark matter. We discuss the main differences between this
class of models and more conventional models of quintessence,
uncoupled to dark matter.}
{}{}

\textheight=7.8truein
\setcounter{footnote}{0}
\renewcommand{\thefootnote}{\alph{footnote}}

\vspace*{0.125truein}

\renewcommand{\theequation}{1.\arabic{equation}}
\setcounter{equation}{0}
\section{Introduction}
\label{sec:1}
\noindent
Understanding Planck-scale physics is one of the main objects of
modern theoretical physics, as an (almost) compulsory ingredient for a
successful unification of all interactions. Present theoretical attempts,
mainly based on supersymmetric models of strings and 
membranes\cite{1},  provide (at least in principle) consistent
unifications schemes including quantized gravitational interactions, at
all energy scales. Their direct verification, however, seems to be out of
reach of the  conventional experimental approach to high energy
physics,  unless we believe in the (probably unlikely) possibility of living
in a ``brane-world" Universe characterized by a very low scale of
(higher-dimensional) bulk gravity\cite{2}. 

Fortunately enough, as discussed  for instance by Starobinski\cite{3}
and Amelino-Camelia\cite{4} also at this Conference,  direct and
important experimental information on quantum gravity and Planck
scale physics is presently coming (or is expected to come soon) from
many astrophysical and cosmological observations. Here, in
particular, I will discuss the possibility of interpreting the large-scale 
acceleration of our present Universe as a direct ``late-time" effect of
dilatonic interactions, correctly described at the Planck scale by string
and M-theory models. 

Let me start by recalling that the string effective action contains, even
to lowest order, at least two fundamental fields, the metric and the
dilaton:
\beq 
S= -{1\over 2 \ls^2} \int d^4x \sqrt{-g} e^{-\phi}
\left[ R+ \left(\nabla \phi\right)^2 + V(\phi) +\dots\right]. 
\label{11}
\eeq
The dilaton $\phi$ is scalar field which, from a physical point of view,
controls the strength of all interactions\cite{5}, in the context of unified
and grand-unified models. From a geometric point of view it may
represent the radius of the $11$-th dimensions\cite{6}, in the context of
M-theory models. Aside from its possible interpretation, the
dilaton is a scalar field necessarily present in the action, it is
non-minimally coupled to the metric and to the other fields, and a 
question which may arise naturally is whether or not such a field can 
automatically provide a model of ``quintessence"\cite{7}, to explain the
cosmic acceleration presently observed on large scales\cite{8}. 

To answer this question we should ask, first of all, what happens to the
dilaton in string cosmology models. Here we shall analyze, in particular,
pre-big bang models\cite{9}, where the dilaton tends to grow\cite{10}
starting from an initial state called ``perturbative vacuum", and 
corresponding to the asymptotic limit $\phi \ra -\infty$. Such a 
growth is not damped, at least initially, by the potential, which is very
flat in the perturbative region (where $V \sim e^{-\exp (-\phi)}$),  so
that the dilaton, sooner or later, necessarily enters the strong coupling
regime. In such a context, {\em large times} are  synonimous of
large values of $\phi$ and {\em large couplings}, and what happens
today thus depends on the form of the dilaton potential in the strong coupling
regime. We have, in principle, two  possibilities. 

A first possibility is that the potential develops some structures
when approaching the strong coupling regime, and that the dilaton
today is frozen, trapped inside a minimum of the potential, in a range
of values which are perturbative enough so as to keep small enough the
effective  coupling $\a_{\rm GUT}$ of grand-unified models (see Fig. 1a).
For instance\cite{11},  $\a_{\rm GUT} \simeq \exp (\phi_0)$, with $\phi_0
\sim -3$. A typical example of potential corresponding to such a
scenario is the following\cite{12}:
\beq
V= m^2 \left[e^{k_1(\phi_-\phi_1)}+ \beta
e^{-k_2(\phi_-\phi_1)}\right] e^{-\ep \exp\left[-\gamma
(\phi_-\phi_1)\right]},
\label{12}
\eeq
where $k_1,k_2, \phi_1,\ep ,\b,\ga$ are
(model-dependent) dimensionless numbers of order one. In such a
context it is possible to obtain realistic cosmological solution\cite{13},
describing a present Universe dominated by the dilaton potential
 and consistent with low-energy gravitational phenomenology.
However, it turns out difficult (if not impossible) to solve the usual
``fine tuning" and ``coincidence" problems (i.e. to explain why the
energy density of the vacuum is so small, and why  it is just of the same order
as the present dark-matter energy density) . 

\begin{figure}[htb]
 \epsfxsize=5cm
 \centerline{\epsfbox{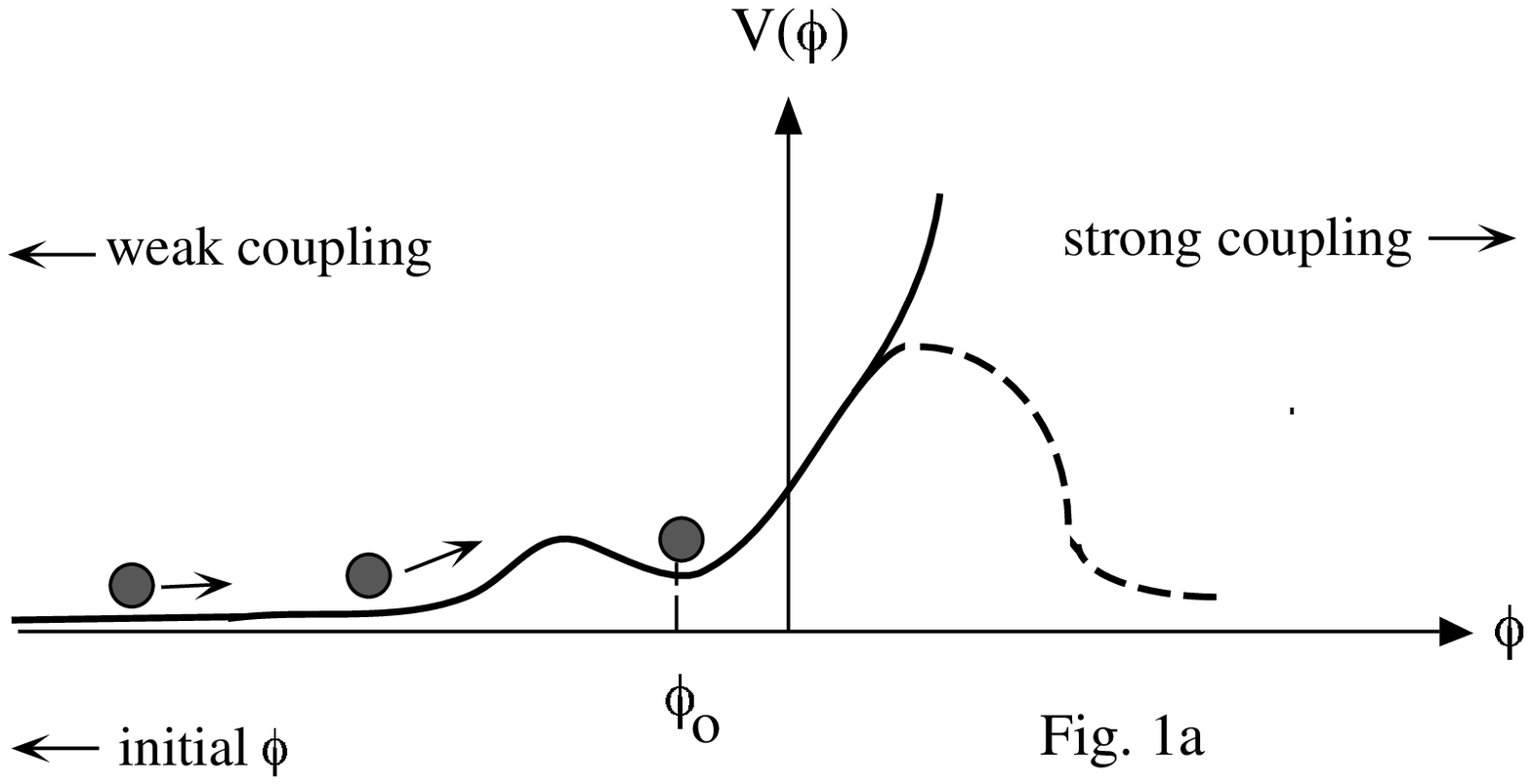} \hskip 1 cm \epsfxsize=5cm
\epsfbox{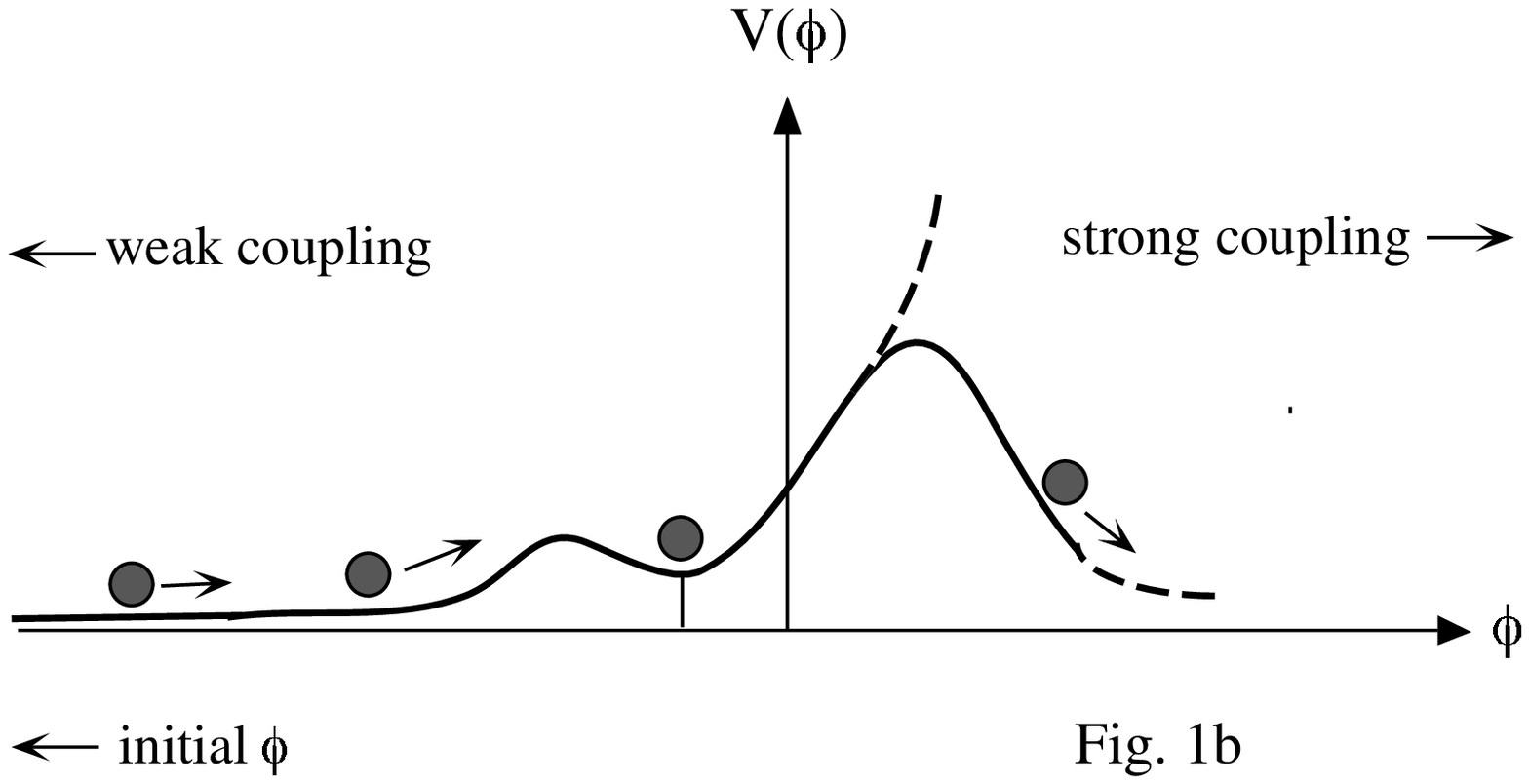}}
\centerline{\parbox{11.5cm}{\caption{\label{fig:f1}
{\sl  Two possible alternative scenarios: the dilaton is trapped in a
semi-perturbative minimum (left), or it is running to plus 
infinity (right).}}}}  
\end{figure}

The second possibility is that the dilaton has not been stopped by the
potential, and that today is free to run to plus infinity, rolling down a 
(probably exponentially) suppressed potential (see Fig. 1b). In that case
the strong coupling corrections to the effective action become more
and more important as time goes on, and one needs an appropriate
``saturation" mechanism to keep the effective couplings small enough,
to be compatible with present phenomenology. For instance\cite{14}:
\beq
\a_{\rm GUT} \simeq e^{\phi_0}\left( 1+N e^{\phi_0}\right)^{-1}, 
~~~~~~~ N\sim 10^2, ~~~~~ ~\phi \ra +\infty, 
\label{13}
\eeq
as will be illustrated in the next section. A typical example of
``bell-like" potential,  corresponding to this scenario, is the
following\cite{15}: 
\beq
V=m^2\left[\exp \left(-e^{-\phi}/\a_1\right)-\exp
\left(-e^{-\phi}/\a_2\right)\right], \label{14}
\eeq
where $\a_1>\a_2>0$, and $c_1$ is a number of order $100$. In such a
context it is possible to obtain realistic cosmological solutions\cite{15}
in which the present Universe is dominated by a mixture of kinetic and
potential dilaton energy density, and it becomes possible to solve --or
at least to relax-- the coincidence problem. The rest of this paper will
be devoted to the discussion on this second possibility. 

\renewcommand{\theequation}{2.\arabic{equation}}
\setcounter{equation}{0}
\section{Late-time saturation of the dilaton couplings}
\label{sec:2}
\noindent
Let us thus consider the large ``bare-coupling" limit, in which $\phi \ra
+\infty$. The string effective action, to lowest order in the
higher-derivative $\ap$ expansion, but to all orders in the dilaton loop
corrections, can be written in this form,  
\beq
S = -\frac{1}{2\,\lambda_s^2}\,\int\,d^{4}x\,\sqrt{-g}\,
\left[e^{-\Psi(\phi)}R+Z(\phi)\partial_{\mu}\phi\partial^{\mu}\phi
+2 \lambda_s^2 V(\phi) \right] + S_m (\phi, g, {\rm matt}),
\label{21}
\eeq
where $\Psi$ ans $Z$ are the  ``form factors" due to the loop
corrections, and other corrections are included into the potential and
the matter action $S_m$. 

In order to obtain the saturation of such corrections we shall follow the
spirit of the old ``induced-gravity" models, assuming the validity of an
asymptotic expansion in inverse powers of  the ``bare"
coupling constant\cite{14,15} $g_s^2= e^\phi$, both for the loop form
factors, for the potential, and for the dilatonic charge $q$ of the matter
fields, defined by $q= -(2/\r_m\sqrt{-g})(\da S_m/\da \phi)$. 
We will set, in particular, 
\bea
&&
e^{-\Psi}=c_1^2+ e^{-\phi} + {\cal O}(e^{-2\phi}), ~~~~~~
Z=-c_2^2+ e^{-\phi} + {\cal O}(e^{-2\phi}), \label{21a} \\
&&
V=V_0 e^{-\phi} + {\cal O}(e^{-2\phi}), ~~~~~~~~~~~~
q=q_0+{\cal O}(e^{-2\phi}).
\label{22}
\eea
Here $c_1^2, c_2^2$ are dimensionless parameters of order one
hundred, as they are controlled by the number of fundamental fields
contributing to the dilatonic loops (this number is large in the context
of unified models). 

In the case of long-range dilaton interactions, that we are
assuming here, we know that the present 
value of the effective dilatonic charge $q(\phi)$ has to be extremely
suppressed for ordinary macroscopic matter ($q\ll1$), to avoid
contradiction with the observed gravitational phenomenology at
low energy. For the (possibly) more exotic components of dark matter, 
on the other hand, there is no need of such suppression,   and the
asymptotic charge $q_0$ could be non-zero, and even large, in principle.
If this is the case we are lead to an interesting and  non-standard
cosmological scenario\cite{15}. 

According to the action (\ref{21}) we must include in fact the dilaton
energy density $\r_\phi$ among the matter sources, in addition to
radiation, baryons and cold dark matter. The Einstein equations (in units
$16 \pi G=1$) can be written in the usual form,   
\beq
6H^2=\r_{rad}+\r_{bar}+\r_{cdm}+\r_{\phi},
\label{23}
\eeq
but the various components of the cosmic fluid are now differently
coupled to the dilaton. Consider, in particular, a model in which the
coupling to ordinary matter decays exponentially with time (i.e.
$q_0=0$ for radiation and baryons), while the coupling to dark matter
grows with time, and tends to saturate to a constant
value $q_0$ as $t \ra \infty$. At late
enough times we then recover the usual conservation equation for
radiation and baryons,
\beq
\dot \r_{rad} +4H \r_{rad}=0, ~~~~~~
\dot \r_{bar} +3H \r_{bar}=0,
\label{24}
\eeq
while the evolution of dark matter is strongly coupled to the dilaton
evolution\cite{16}, through a model-dependent coupling
function $\ep(\phi)$, 
\beq
\dot \r_{cdm} +3H \r_{cdm} -{1\over 2} \ep(\phi) \hpd \r_{cdm}=0,
~~~~~
{\ddot {\hp}} +3H \hpd + {\pa V\over \pa \hp} +{1\over 2} \ep(\phi)
\r_{cdm}=0. 
\label{25}
\eeq
Here $\hp$ is the canonically rescaled dilaton field in the
Einstein frame, and $\ep(\phi)$ is determined\cite{15,17} by the explicit
form of the asymptotic expansion (\ref{22}). 

Because of this coupling, there are possible drastic modifications of the
standard scenario after equilibrium. When the Universe become
matter-dominated, the dilaton (still subdominant) is
``dragged" by dark matter, and we may have a first (very slight)
modification of the standard decelerated evolution, since $\r_{cdm}
\sim \r_\phi \sim a^{-(3+\ep^2)}$, where $a$ is the scale factor 
(of course, $\ep<1$ to avoid contradictions with standard gravitational
phenomenology). When the dilaton potential comes into play,
eventually, the Universe enters a final phase in
which $\r_{cdm} \sim \r_\phi \sim V \sim a^{-6/(2+q_0)} \sim H^2$, 
which is accelerated for  $q_0>1$, and in which the ratio of the dilaton
to dark-matter energy density is frozen, asymptotically, to a constant
number of order one (depending on $q_0,c_1,c_2$). The fraction of
kinetic to potential energy density of the dilaton (i.e., the equation of
state of the dark energy component) is also fixed by the
parameters of the asymptotic expansion (\ref{22}).

To give a qualitative illustration of such scenario we present in Fig. 2
a plot of the various energy densities, as a function of $a$ and of the
red-shift parameter $z=a_0/a-1$. The curves have been 
obtained through a numerical integration of the string
cosmology equations\cite{15} obtained from the action (\ref{21}),
with the potential (\ref{14}) and with $\Psi,Z$ defined by the expansion
(\ref{21a}) truncated to first order. 
We have set  $q(\phi)=q_0 e^{q_0
\phi}/(c^2 + e^{q_0 \phi})$, $q_0=2.5$, $c^2=150$,  $c_1^2=100$,
$c_2^2=30$, $\a_1=10=2\a_2$. Finally, we have appropriately tuned the
potential by choosing $m= 10^{-3}H_{\rm eq}$,  so that today (i.e. for
$z=0$) we are already inside the dilaton-dominated evolution. In the
radiation phase we can see the existence of  a ``focusing effect",  by
which the (subdominant) dilaton energy tends to approach the other
components.  In the dragging phase the dilaton kinetic energy closely
follows the dark-matter evolution. In the final freezing phase the
evolution of the dilaton and dark-matter energy densities are closely
tied toghether, and their ratio is fixed for ever to a number of order one. 

\begin{figure}[htb]
 \epsfxsize=8cm
 \centerline{\epsfbox{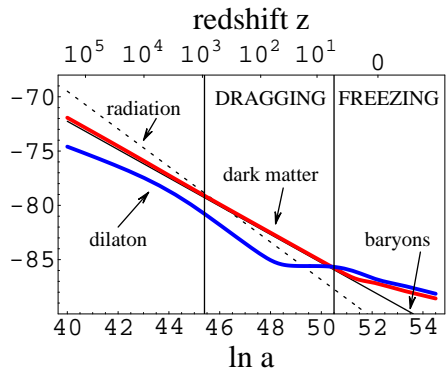}}
\centerline{\parbox{11.5cm}{\caption{\label{fig:f2}
{\sl  Time evolution of the various energy densities. Radiation and
baryons are uncoupled to the dilaton, and thus obey the standard
scaling behaviour ($\r\sim a^{-4}$ and $\r \sim a^{-3}$,
respectively).}}}}    
\end{figure}

\renewcommand{\theequation}{3.\arabic{equation}}
\setcounter{equation}{0}
\section{Phenomenological consequences of a ``running-dilaton"
cosmology} 
\label{sec:3}
\noindent
There are three main differences between the dilatonic models of dark
energy (coupled to dark matter) illustrated before, and the  more
conventional  (uncoupled) models of dark energy. 

$\bullet$ The first difference, also evident from Fig. 2, is that  in the
final  freezing phase the energy density of dark matter is diluted in time
at a slower  ratio than the energy density of baryons, which are
uncoupled to the dilaton. As a consequence, the ratio
$\r_{bar}/\r_{cdm}\sim a^{-3 q_0/(2+q_0)}$   {\em decreases} in time
during the accelerated phase, and this could probably explain why today
the fraction of baryons is so small ($\sim 10^{-2}$) in critical units.
Experimental information on the past value of the ratio
$\r_{bar}/\r_{cdm}$, if available, and if compared with the present value
of this ratio, would immediately provide a direct test of this class of
dilatonic models. 

$\bullet$ The second difference is that the coincidence problem, if not
solved, is  at least relaxed, because the dilatonic (dark-energy) density
and the  dark-matter density are of the same order {\em not only 
today}, but  also in the future (for ever), and possibly in the past,
depending on the beginning of the freezing epoch (i.e., on the specific
amplitude $V_0$ of the potential). 

$\bullet$ The possibility of an early beginning of the accelerated
epoch\cite{17}, even at redshift values $z_{acc}$ much larger than one,
is indeed the third  important difference characterizing our class of
dilatonic models. In models of uncoupled dark energy, indeed, 
$z_{acc}$ is always smaller than one, or at most  one. 

For a more detailed illustration of this important point consider in
fact the Einstein equations, for a model of uncoupled quintessence with
energy density $\r_Q$,
\beq
6H^2= \r_{m}+\r_Q, ~~~~~~~~
4\dot H +6H^2 =-p_Q=-w ~\r_Q,
\eeq
and with fixed eqation of state, 
\beq
-1 \laq p_Q/\r_Q \equiv w \laq -1/3
\label{32}
\eeq
(here $\r_m$ includes both baryons and dark matter). 
In such a context, $\r_{m}\sim a^{-3}$ is diluted faster than $\r_Q
\sim a^{-3(1+w)}$. So, even if today $\r_Q$ dominates, and the expansion
is accelerated ($\ddot a/a= \dot H+H^2>0$), at some time in the past
$\r_{m}$ necessarily has to become dominant, in a decelerated
Universe. The acceleration, in particular, switches off ($\ddot a=0$) at 
\beq
z_{acc}\equiv {a_0\over
a_{acc}}-1=\left[(1+3w)\left(\Om_{m}-1\over
\Om_{m}\right)\right]^{-1/3w} -1,
\label{33}
\eeq
where $\Om_m =\r_m/6H^2$. If we plot this function $z_{acc}(w)$, at 
(fixed) realistic values of $\Om_m$,  we can easily check that $z_{acc}
\laq 1$ for realistic (i.e. observationally compatible) values of the 
dark-energy equation of state (in the range of eq. (\ref{32})), as 
illustrated in Fig. 3. 

\begin{figure}[htb]
 \epsfxsize=6cm
 \centerline{\epsfbox{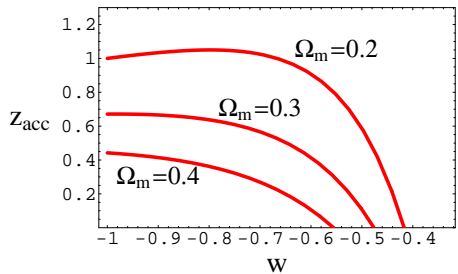}}
\centerline{\parbox{11.5cm}{\caption{\label{fig:f3}
{\sl  Possible beginning of the accelerated epoch for models of
uncoupled dark energy with fixed equation of state. }}}}     
\end{figure}

The asymptotic freezing phase of dilatonic models, on the contrary, is
characterized by a constant positive acceleration\cite{15} $\ddot
a/(aH^2)=(q_0-1)/(q_0+2)>0)$, and its extension towards the past is
in principle constrained by the present fraction of baryon energy
density, which grows as we go back in time, in such a way that baryons
tend to become over-dominant. By imposing that this is not the case, we
can obtain more significant constraints on $z_{acc}$ from the measured
value of the so-colled density contrast $\sg_8$, characterizing the
level of dark-matter fluctuations over a distance scale $R_8 \simeq 8$
Mpc. 

We have used the current SNIa observations\cite{18} to extract
information on the parameters $c_1$, $c_2$, $q_0$ of the freezing
phase (determining the asymptotic value of $w$), and we have fixed
the present values of $\r_{cdm}$, $\r_{bar}$ to the standard values
$\Om_{cdm}=0.3$, $\Om_{bar}h^2=0.02$, $h=0.65$. By imposing the
$\sg_8$ constraint on the density fluctuations predicted in models of
coupled dark energy\cite{19}, one then finds\cite{17} that the
beginning of the dilaton-dominated epoch is allowed up to $z_{acc}
\simeq 3.5$, according to the best fit value of $w$ from present SNIa
data; taking into account one sigma and two sigma deviations
from the best fit one can also obtain, respectively, $z_{acc}\laq 5$ and
$z_{acc} \laq 8$. In any case, the possible past extension of the
accelerated phase is considerably increased with respect to the
standard predictions of Fig. 3, thus providing a further relaxation of the
coincidence problem. 

It is finally worth stressing that such an early beginning of the
acceleration is compatible not only with $\sg_8$ measurements, but also
with the  recent observations of the farthest Supernova SN1997ff at
$z=1.7$. This point is illustrated in Fig. 4, where we have reported the
distance-modulus versus the redshift,  for all the high-redshift
Supernovae known so far. The dashed curve corresponds to the
luminosity-distance relation for a standard $\La CDM$ model, which is
in good agreement with all data, but which is accelerated only up to
$z\simeq 0.5$. The full curves correspond instead to dilatonic dark
energy models. The bold ones also include baryons, which become more
and more important as we go back in time. We have plotted, particular,
two curves, corresponding to the best fit ($\b_2=4.02$) and to the
one-sigma deviation ($\b_2=2.35$) of the low-redshift data. Both curves
are also in agreement with the Supernova at $z=1.7$, in spite of the fact
that the acceleration, for the associated cosmological model,  starts
well above $z=1.7$. 

\begin{figure}[htb]
 \epsfxsize=10cm
 \centerline{\epsfbox{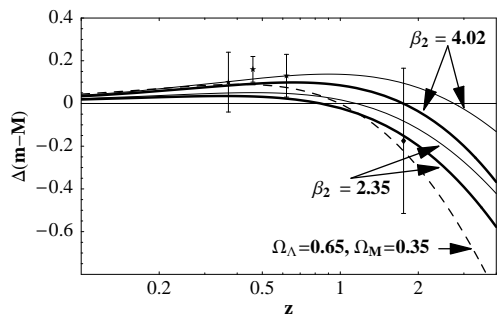}}
\centerline{\parbox{11.5cm}{\caption{\label{fig:f4}
{\sl  The farthest Supernova at $z=1.7$ is compatible both with an early
decelerated Universe (in the context of uncoupled models of dark
energy), and with an early accelerated Universe (in the context of the
dilatonic models discussed in this paper).}}}}     
\end{figure}

\renewcommand{\theequation}{5.\arabic{equation}}
\setcounter{equation}{0}
\section{Concluding remarks}
\label{sec:5}
\noindent
In the context of string cosmology it is possible to formulate
consistent models in which our present accelerated Universe is
dominated by a mixture of kinetic plus potential dilaton energy-density.
This requires that the dilaton loop corrections are
asymptotically saturated (to keep small enough the effective couplings), 
and non-universal (a strong coupling to dark matter is required, in
particular). When the above assumptions are satisfied it can be shown,
in addition, that  the approximate equality of dark-matter and
dark-energy density is no longer a coincidence typical of the present
epoch.

There are two important phenomenological segnatures of such a class
of dilatonic dark-energy models: 1) the time variation of the ratio
$\r_{bar}/\r_{cdm}$ during the accelerated phase; 2) the possibility of
an early beginning of the acceleration, well above $z=1$. A direct 
test of this second prediction is possibly expected from work in
progress on the luminosity-redshift distributions of gamma-ray
bursts\cite{21}, using their sources as standard candles covering a
range of redshift-values much larger than in the case of Supernovae
observations. 

\vspace{0.5cm}
\noi
{\it Acknowledgments:\/} I am very grateful to  Luca
Amendola, Federico Piazza, Domenico Tocchini-Valentini, Carlo Ungarelli
and Gabriele Veneziano for the pleasant and fruitful
collaboration leading to the results presented in this paper.

\end{document}